\documentclass[10pt,letterpaper]{article}
\usepackage{graphicx}
\usepackage{amsmath}
\usepackage{verbatim} 
\usepackage{color}
\usepackage{opex3}

\begin{document}

\title{Optical coupling to nanoscale optomechanical cavities for near quantum-limited motion transduction}
\author{Justin D. Cohen$^\dag$, Se\'{a}n M. Meenehan$^\dag$, and Oskar Painter$^{*}$}
\address{Kavli Nanoscience Institute and Thomas J. Watson, Sr. Laboratory of Applied Physics, California Institute of Technology, Pasadena, CA 91125, USA. \\
Institute for Quantum Information and Matter, California Institute of Technology, Pasadena, CA 91125, USA.}
\email{$^{*}$opainter@caltech.edu}

\begin{abstract}
A significant challenge in the development of chip-scale cavity-optomechanical devices as testbeds for quantum experiments and classical metrology lies in the coupling of light from nanoscale optical mode volumes to conventional optical components such as lenses and fibers. In this work we demonstrate a high-efficiency, single-sided fiber-optic coupling platform for optomechanical cavities. By utilizing an adiabatic waveguide taper to transform a single optical mode between a photonic crystal zipper cavity and a permanently mounted fiber, we achieve a collection efficiency for intracavity photons of $52 \%$ at the cavity resonance wavelength of $\lambda \approx 1538$ nm. An optical balanced homodyne measurement of the displacement fluctuations of the fundamental in-plane mechanical resonance at $3.3$ MHz reveals that the imprecision noise floor lies a factor of $2.8$ above the standard quantum limit (SQL) for continuous position measurement, with a predicted total added noise of $1.4$ phonons at the optimal probe power. The combination of extremely low measurement noise and robust fiber alignment presents significant progress towards single-phonon sensitivity for these sorts of integrated micro-optomechanical cavities.
\end{abstract}

\ocis{(230.5298)~Photonic crystals; (230.3120)~Integrated optics devices; (220.4880)~Optomechanics; (280.4788)~Optical sensing and sensors; (350.4238)~Nanophotonics and photonic crystals}

\bibliographystyle{OSA}

\section{Introduction}
Nanoscale structures in the form of photonic and phononic crystals have recently been shown to feature significant radiation pressure interactions between localized optical cavity modes and internal nanomechanical resonances~\cite{Eichenfield20092}. Alongside similar advances in the microwave domain~\cite{Teufel2011}, optomechanical crystals have recently been used to laser cool a nanomechanical oscillator to its quantum ground state~\cite{Chan2011}. The ability to measure and control the quantum state of such an object ultimately hinges on the quantum efficiency of the optical transduction of motion. Here we demonstrate high-efficiency optical coupling between an optomechanical zipper cavity~\cite{Eichenfield2009} and a permanently mounted optical fiber through adiabatic mode conversion. This optical coupling technique greatly improves the collection efficiency of light from these types of optomechanical cavities over existing methods, and brings the minimum total added measurement noise to within a factor of $3$ of the standard-quantum-limit of continuous position measurement.

In a weak measurement of position through a parametrically coupled optical cavity there are two intrinsic sources of measurement noise. Shot noise of the probe laser and excess quantum vacuum noise due to optical signal loss set the fundamental noise floor of the measurement. When converted into units of mechanical quanta this imprecision noise $N_{\text{imp}}$ decreases with increasing probe power. However, higher laser probe power comes at the cost of radiation pressure backaction driving an additional occupation noise $N_{\text{BA}}$ on top of the thermal mode occupation $N_{\text{th}}$. For an optically resonant measurement of position, the noise terms in units of mechanical occupation quanta are
\begin{equation} \label{eqn:Noise}
N_{\text{imp}} = \frac{\kappa^2 \gamma}{64 n_c g^2 \kappa_e \eta_{\text{cpl}}\eta_{\text{meas}}}, \quad
N_{\text{BA}} = \frac{4 n_c g^2}{\kappa \gamma}, 
\end{equation}
\noindent where $g$ is the optomechanical interaction rate, $\kappa$ and $\gamma$ are respectively the optical and mechanical loss rates, $\kappa_e$ is the extrinsic cavity loss rate, $\eta_{\text{cpl}}$ is the optical efficiency between the cavity and the detection channel, and $\eta_{\text{meas}}$ accounts for excess technical noise and signal loss accumulated in experiment-specific optical components. As intracavity photon number $n_c$ is varied, an optimal input power $P_{\text{min}}$ is reached where the imprecision noise and back-action noises are equal and the total added noise is minimized to
\begin{equation} \label{eqn:NoiseOpt}
N_{\text{min}}=\left(N_{\text{imp}} + N_{\text{BA}} \right )_{\text{min}} = \frac{1}{2 \sqrt{\eta_{\text{cpl}}\eta_{\text{meas}} \kappa_e/\kappa}}.
\end{equation}

In the ideal case this point, known as the standard-quantum-limit (SQL), adds $1/2$ quanta of noise to the measurement, equal to the zero-point fluctuations of the oscillator~\cite{Clerk2010}. The SQL can only be reached in the limit of noise-free, lossless detection ($\eta_{\text{cpl}}\eta_{\text{meas}} = 1$) and perfect waveguide loading ($\kappa_e = \kappa$). Thus, $N_{\text{min}}$ is a suitable figure of merit for the ultimate quantum efficiency of an optomechanical detector of position. Although experiments in both the optical and microwave domains have brought the imprecision noise level down to below $1/4$ quanta~\cite{Teufel2009,Anetsberger2010}, and recent microwave experiments have acheived total added noise within a factor of $4$ of the SQL~\cite{Teufel2011}, current state-of-the-art optical devices have been limited to $14-80$ times the SQL~\cite{Chan2011,Riviere2011}. Such experiments are limited partly by technical noise (e.g. added noise from amplifiers), but a substantial amount of imprecision is introduced by poor quantum efficiency of the optical readout. Here we identify another figure of merit to allow for cross-platform comparison of detection methods. The quantum efficiency of a general measurement apparatus will be limited to $\eta_{\text{CE}}=\eta_{\text{cpl}}\kappa_e/\kappa$, the collection efficiency of intracavity photons into the detection channel before further signal processing.

To date, most nanoscale optomechanical experiments utilize evanescent coupling between the optical cavity and an adiabatically tapered optical fiber~\cite{Michael2007,Spillane2003}. While this method offers low parasitic losses, standing wave resonators such as the optomechanical cavities considered here radiate symmetrically into two oppositely propagating modes of the fiber, each at a rate $\kappa_e$. Thus the fraction of $n_c$ routed into the detection channel $\kappa_e/\kappa = \kappa_e/(\kappa_i+2\kappa_e)$  does not exceed $1/2$ even in the ideal case of negligible intrinsic cavity loss rate $\kappa_i$. To reach the overcoupled regime $\kappa_e/\kappa > 1/2$, a single-sided coupling scheme is necessary.

\begin{figure*}
\includegraphics[width=\columnwidth]{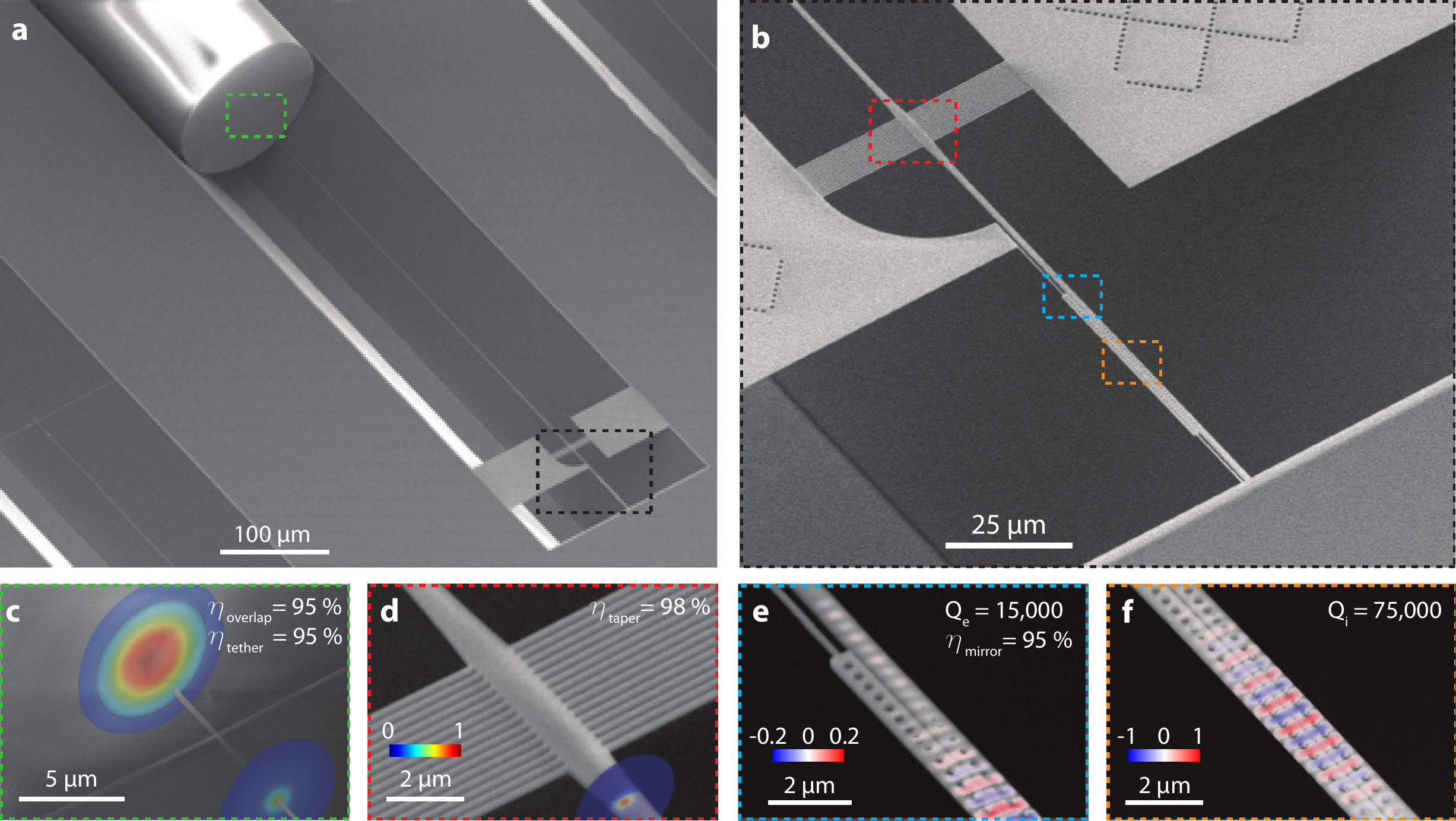}
\caption[width=\columnwidth]{Scanning electron microscope (SEM) images illustrating the optical coupling scheme and mode conversion junctions, with overlayed mode profiles simulated via Finite-Element-Method (FEM) of optical power in (c) and (d), and electric field in (e) and (f). (a) Fabricated device after fiber coupling via self-aligned v-groove placement. (b) Detailed view of the zipper cavity. (c) The optical-fiber/Si$_3$N$_4$-waveguide junction. (d) The waveguide with supporting tethers after adiabatically widening to $1.5$ $\mu$m. (e) Photonic crystal taper section. (f) Photonic crystal defect cavity.}
\label{fig:fig_1}
\end{figure*}

\section{Single-sided Fiber Coupling}
We have implemented a coupling scheme which routes light from a photonic crystal zipper cavity to a cleaved single-mode optical fiber tip with high efficiency in a fully single-sided interface. The fiber is self-aligned in a Si V-groove (Fig. \ref{fig:fig_1}), secured with epoxy, and butt-coupled to a mode-matched Si$_3$N$_4$ waveguide, which adiabatically widens~\cite{Mitomi1994,Almeida2003,McNab2003,Chen2010} to match the width of the zipper cavity nanobeam. The waveguide then couples to the cavity through a truncated photonic crystal mirror. The robust fiber alignment offers another key advantage over evanescent and grating-coupler techniques that call for nanometer-scale-sensitive positioning to achieve appreciable mode overlap. Nano-positioning is difficult and expensive to implement in cryogenic setups due to footprint and imaging requirements, whereas the coupler presented here can be installed directly into any system with a fiber port.

We now describe the optimization of the optical efficiency. To determine the optimal width of the Si$_3$N$_4$ waveguide, we compute the guided transverse modes of both the waveguide and the optical fiber at the target wavelength of $\lambda = 1550$ nm using a finite-element-method (FEM) solver~\cite{COMSOL}. The coupling efficiency at the fiber-waveguide junction is calculated from the mode profiles using a mode overlap integral~\cite{SnyderLove}. For a $400$ nm thick Si$_3$N$_4$ membrane, the optimal waveguide width is $w=230$ nm with a transmission efficiency of $\eta_{\text{overlap}} = 95 \%$ as depicted in Fig.  \ref{fig:fig_1}c. Then $w$  increases gradually to the nanobeam width of $850$ nm. To obtain high transmission efficiency in this tapered waveguide section, the rate of change of $w$ along the propagation direction $z$ must be small enough to satisfy the adiabatic condition~\cite{SnyderLove} $dw/dz \ll \Delta n_{\text{eff}}$ at every point along the taper, where $\Delta n_{\text{eff}}$ is the difference in effective index between the fundamental waveguide mode and any other guided or radiation mode. The actual transmission efficiency is calculated using a finite-difference-time-domain (FDTD) simulation~\cite{Lumerical}. For a $400 \mu$m long taper with a cubic shape between the junctions shown in Fig.  \ref{fig:fig_1}c,d, we obtain a transmission efficiency of $\eta_{\text{taper}} = 98 \%$.

The tapered waveguide is supported by a $70$ nm wide tether placed near the fiber-waveguide junction (Fig.  \ref{fig:fig_1}c). The scattering loss of the tether is computed using FDTD and the transmission efficiency is calculated to be $\eta_{\text{tether}} = 95 \%$. A second set of $150$ nm wide tethers (Fig.  \ref{fig:fig_1}d) is placed just before the cavity, in order to isolate the optomechanical crystal from the low-frequency vibrational modes of the tapered waveguide. The waveguide is temporarily widened to $1.5\; \mu$m at this point, rendering the scattering loss due to the tethers negligible.

Finally, the uniform dielectric waveguide adiabatically transitions into a one-dimensional photonic crystal mirror by linearly increasing the radius of the holes while keeping the lattice constant fixed. An $8$ hole photonic crystal taper (Fig \ref{fig:fig_1}e) is sufficient to acheive an efficiency of $\eta_{\text{mirror}} = 95 \%$. The coupling rate to the cavity is controlled by varying the number of mirror periods after the taper. The photonic crystal cavity mode (Fig. \ref{fig:fig_1}f) is shared between the waveguide beam and a near-field, flexibly supported test beam, with the optomechanical coupling arising from the sensitivity of the resonance frequency to the beam separation. In this manner the waveguide structure serves as an optical readout of the test beam motion with an optimal round-trip efficiency of $\eta_{\text{rt}} = (\eta_{\text{overlap}}\eta_{\text{taper}} \eta_{\text{tether}})^2 \eta_{\text{mirror}} = 74 \%$. For the purposes of optomechanical transduction, the relevant figure is the single-pass transmission efficiency between the cavity output and the fiber, which ideally is $\eta_{\text{cpl}} = \sqrt{\eta_{\text{rt}}} = 86 \%$.

\begin{figure*}
\includegraphics[width=\columnwidth]{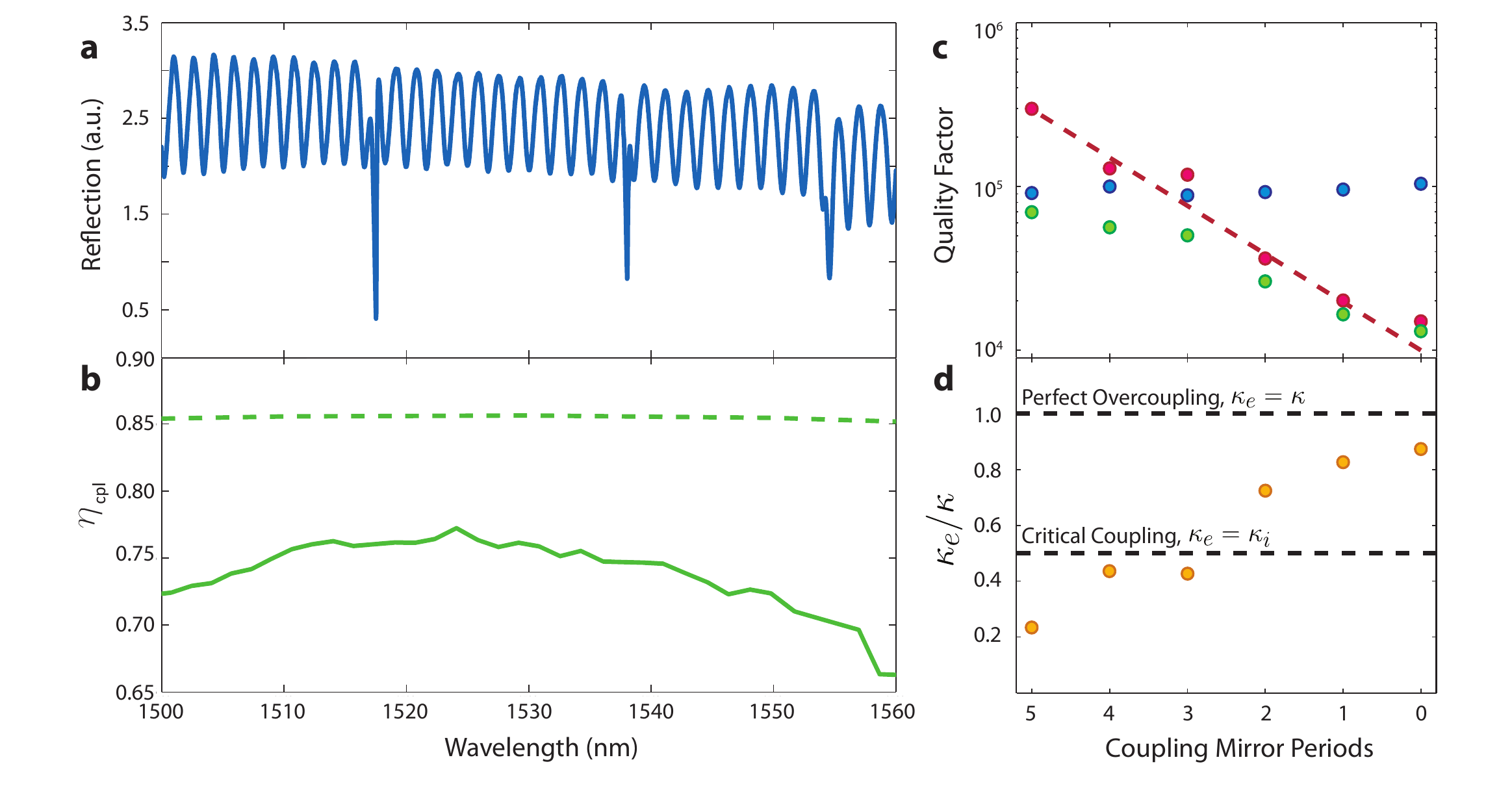}
\caption[width=\columnwidth]{Optical response of the system. (a) A wide range reflection scan reveals Fabry-Perot interference fringes off-resonant from the photonic crystal cavity, and sharp dips on resonance. The visibility of the fringes reveals that $\eta_{\text{cpl}}$ has the wavelength dependence shown in solid green in (b), with $\eta_{\text{cpl}}=74.6\;\%$ at the cavity resonance wavelength of $\lambda \approx 1538$ nm. The dashed green line denotes the simulated ideal efficiency of the coupler. (c) By tuning the number of mirror-type-hole periods between the coupling section and cavity section of the photonic crystal, the total quality factor $Q_t$ (green points) of the resonance transitions from being limited by intrinsic loss $Q_i$ (blue points) to extrinsic loss $Q_e$ (red points) in good agreement with simulation (dashed red curve). (d) The ratio of extrinsic loss rate $\kappa_e$ to total loss rate $\kappa$ progresses from the strongly undercoupled regime to the strongly overcoupled regime with the mirror variation.}
\label{fig:fig_2}
\end{figure*}

\section{Optical Characterization}
A broadband scan of the reflection from the fiber-coupled device is shown in Fig. \ref{fig:fig_2}a. For probe wavelengths detuned from cavity resonances, the photonic crystal depicted in Fig. \ref{fig:fig_1}d,e functions as a near-unity reflectivity mirror. Thus a low-finesse Fabry-Perot cavity is formed in the waveguide between the photonic crystal and the cleaved fiber facet (with reflectivity $R = 3.5 \%$) of Fig \ref{fig:fig_1}b. By fitting the visibility of the fringes to $V = \eta_{\text{cpl}}(1 - R)/(\sqrt{R}(1 - \eta_{\text{cpl}}^2))$, the curve in Fig. \ref{fig:fig_2}a provides a convenient calibration of optical efficiency, which is plotted in Fig. \ref{fig:fig_2}b versus wavelength and reveals $\eta_{\text{cpl}} = 74.6\%$ for resonant measurements of the primary mode at $\lambda = 1538$ nm. The coupling depth of the mode is determined by fitting the resonance dip to a coupled-cavity model incorporating the photonic crystal and Fabry-Perot interference effects, yielding  $\kappa_e/\kappa = 0.7$. To verify this model, we study a series of devices with varying numbers of mirror-type holes between the photonic crystal taper and cavity. In Fig. \ref{fig:fig_2}c, the total optical quality factor ($Q_t$) is plotted in green alongside intrinsic ($Q_i$) and extrinsic ($Q_e$) components as determined by the coupled cavity fit (blue and red points respectively). As the mirror holes are removed, $Q_e$ decreases in good agreement with simulation (dotted line), and the limiting component of $Q_t$ transitions from $Q_i$ to $Q_e$. By converting quality factor into loss rate and calculating the coupling depth of the device series, a clear trend from undercoupling to overcoupling is evident in Fig. \ref{fig:fig_2}d. The device under test in this work features $\kappa_e/\kappa = 0.7$, bringing the fiber collection efficiency figure of merit to $\eta_{\text{CE}} = 52 \%$. For comparison, systems in which ground-state occupancy of a single mechanical mode has been achieved have featured $\eta_{\text{CE}} = 37\; \%$~\cite{Teufel2011} and $\eta_{\text{CE}} = 9\; \%$~\cite{Chan2011}.

\begin{figure*}
\includegraphics[width=\columnwidth]{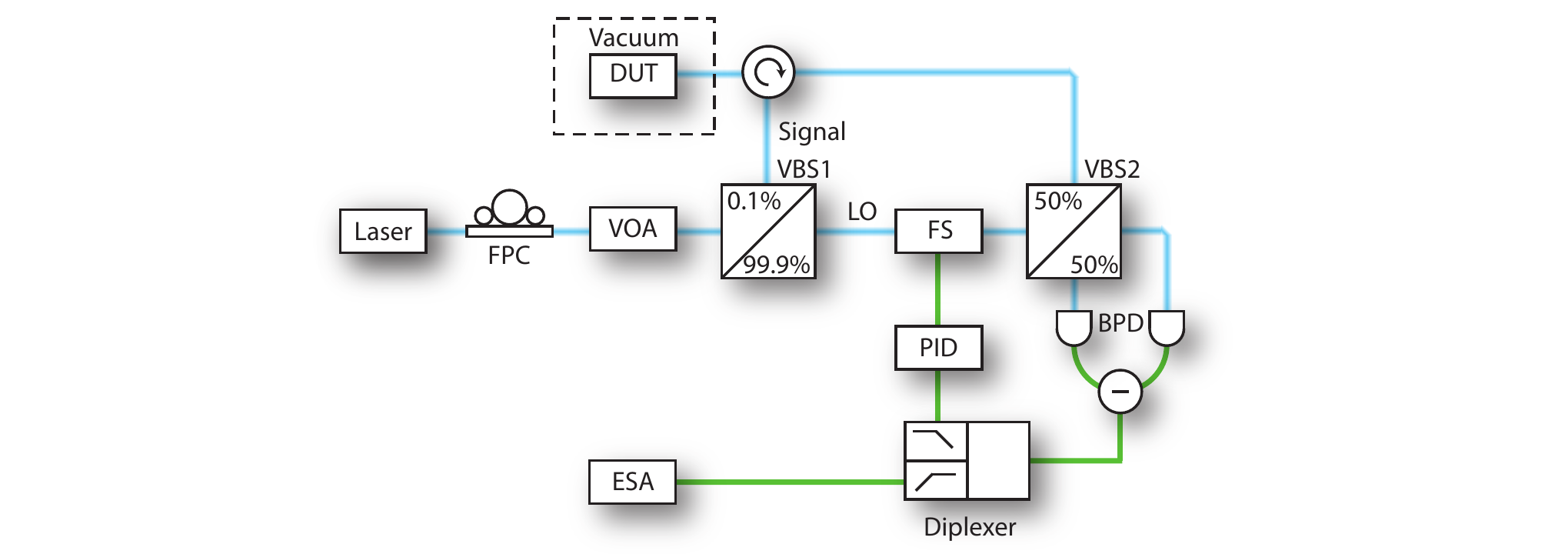}
\caption[width=\columnwidth]{Experimental setup. FPC: fiber polarization controller. VOA: variable optical attenuator. VBS: variable beam splitter. LO: local oscillator. DUT: device under test. FS: fiber stretcher. PID: proportional-integral-derivative controller. BPD: balanced photodiode. ESA: electronic spectrum analyzer.}
\label{fig:fig_3}
\end{figure*}

\section{Mechanical Spectroscopy}
\begin{figure*}
\includegraphics[width=\columnwidth]{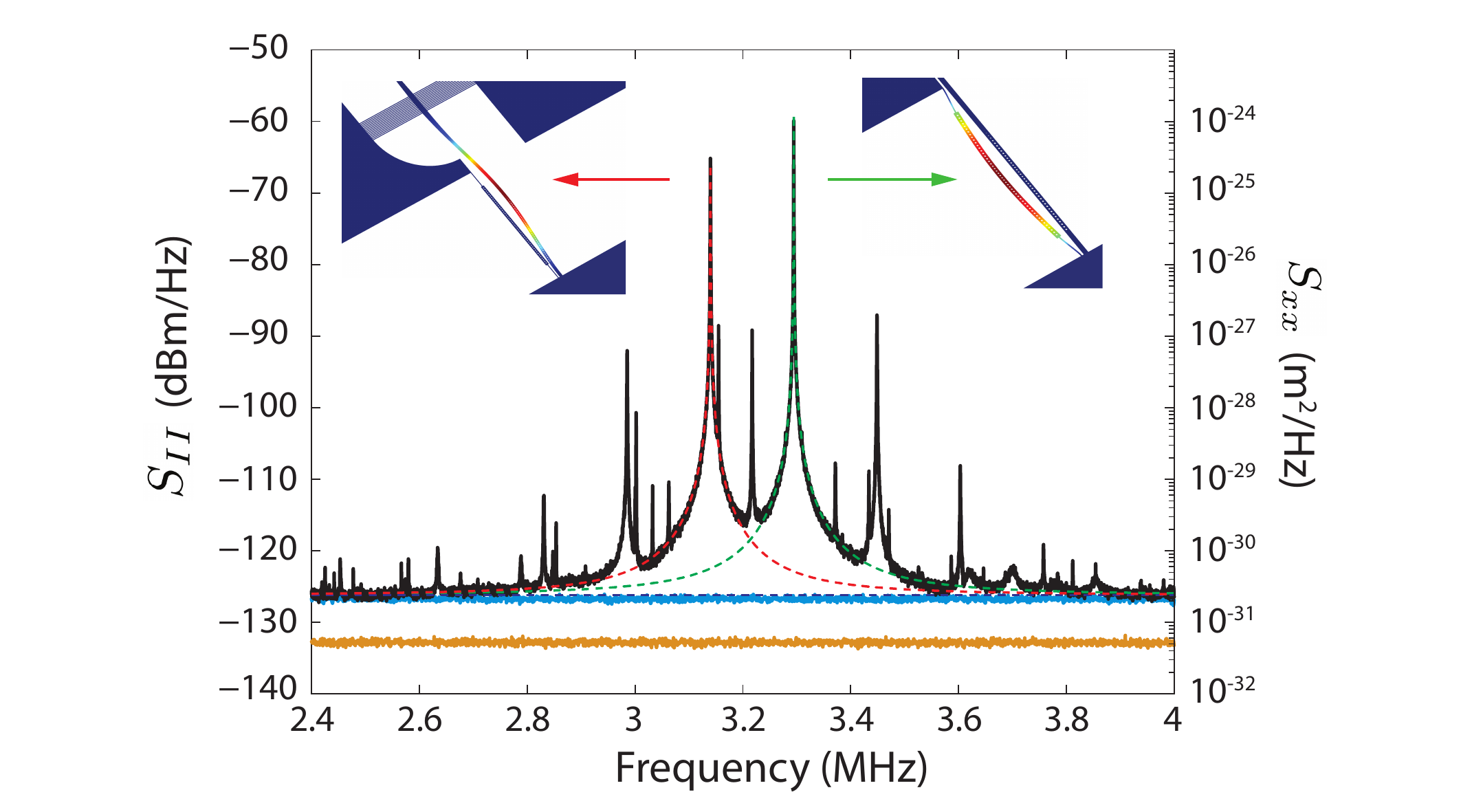}
\caption[width=\columnwidth]{Photocurrent NPSD $S_{II}(\omega)$ (solid black curve) measured on the ESA with $10$ nW of input signal arm power. Optical shot noise (solid blue curve) sets the noise floor several dBm above electronic noise contributions (solid orange curve.) The two peaks at $3.14$ MHz and $3.3$ MHz correspond to the first order mechanical bending modes of the waveguide and test beams respectively. The dashed red and green curves display the calculated single-sided displacement NPSD of each mode, with the calculated imprecision noise floor shown in the dashed purple line. (insets) FEM simulated mechanical bending modes of each beam.}
\label{fig:fig_4}
\end{figure*}

To extract the noise power spectral density (NPSD) of the test-beam displacement $S_{xx}$ from the optomechanically phase-shifted cavity reflection, we place the device in an evacuated environment and connect the output fiber to the signal arm of a fiber-based balanced homodyne receiver shown schematically in Fig \ref{fig:fig_3}. We resonantly probe the optical resonance to avoid dynamic back-action on the test beam as well as to maximize the mechanical transduction sensitivity. The probe laser frequency is positioned on resonance without frequency-locking to the cavity as drift is found to be negligibile over the time span of the measurement ($\sim 100$ s). The laser output is sent through a fiber polarization controller (FPC) and variable optical attenuator (VOA) before being split by a variable beam splitter (VBS1) into into the signal and local oscillator (LO) arms of the homodyne receiver. The signal circulates through the device, while the LO propagates through a fiber stretcher (FS) before recombining with the signal at VBS2, with the interference output detected on a balanced photodiode (BPD) pair. The difference signal generated by the BPD is dediplexed at $200$ kHz into a low-frequency lock signal (fed back through a proportional-integral-derivative (PID) controller onto the FS to eliminate path-length fluctuations in the interferometer) and a high-frequency photocurrent response. The NPSD of this high-frequency signal, $S_{II}(\omega)$, is measured on an electronic spectrum analyzer (ESA) and plotted in Fig. \ref{fig:fig_4} for a probe power of $10$ nW. The optical efficiency of the setup to the device reflection signal is set by the transmission through the circulator ($87\;\%$) and VBS2 ($94\;\%$) as well as the BPD quantum efficiency ($44\;\%$), which combine to $\eta_{\text{meas}}=36\%$.

The NPSD shown in Fig. \ref{fig:fig_4} reveals a prominent resonance at  $\omega=3.3$ MHz, which is identified using FEM simulations as the fundamental in-plane mechanical mode of the test beam. The optomechanical coupling rate of this mode, calibrated through the optical spring shift~\cite{Eichenfield2009}, is $g/2\pi=350$ kHz. A second prominent resonance exists at $\omega=3.15$ MHz which is identified as the fundamental in-plane mechanical mode of the waveguide with a lower optomechanical coupling ($g/2\pi=135$ kHz) owing to small overlap between the mechanical and optical modes. Overlays of the predicted single-sided transduction spectra, calculated using the measured parameters of the optomechanical cavity and fiber coupler, show good agreement with the measured signal for both the test-beam (dashed green curve) and the waveguide (dashed red curve). Other peaks at $3$ MHz, $3.2$ MHz, and $3.4$ MHz do not correspond to any real mechanical motion of the beam but rather are due to nonlinear transduction of the mechanics. Thermal Brownian motion of the beams, combined with the large optomechanical coupling, gives rise to a frequency shift of the optical mode which is a substantial fraction of the cavity optical linewidth, leading to harmonics in $S_{II}(\omega)$ at multiples of the sum and difference frequencies of the two mechanical modes~\cite{Lin2009}. The remaining small features in $S_{II}(\omega)$ are due to out-of-plane flexural modes of the structure, which are weakly transduced due to imperfect vertical symmetry.

\section{Measurement Imprecision and Discussion}
For the remainder of the paper we focus our discussion on measurement of the mechanical mode with the strongest transduction, the fundamental in-plane mode of the test beam with mechanical frequency $\omega_\text{m} = 3.3$ MHz. The single-sided NPSD of displacement for a harmonic mechanical mode at the mechanical resonance frequency is given by $S_{xx}(\omega_m) = 8 x_{\text{ZPF}}^2 (\langle n \rangle+1/2)/\gamma$, where $x_{\text{ZPF}} = \sqrt{\hbar/2 m \omega_\text{m}}$ is the zero-point amplitude of the mechanical resonator with effective motional mass $m$, and $\langle n \rangle = N_{\text{th}} + N_{\text{BA}}$ is the mode occupancy in units of phonon quanta. For the fundamental in-plane mode we numerically compute an effective mass of $m = 15$ pg and a corresponding zero-point amplitude of $x_{\text{ZPF}} = 13$ fm. When the mechanical mode is well-resolved in $S_{II}(\omega)$, that is, when the contributions of nearby mechanical modes are negligible and the resolution bandwidth of the ESA is much less than $\gamma$, we can convert the spectrum into units of displacement for the measured mechanical mode by scaling $S_{II}(\omega_\text{m})$ to the computed value of $S_{xx}(\omega_\text{m})$.

\begin{figure*}
\includegraphics[width=\columnwidth]{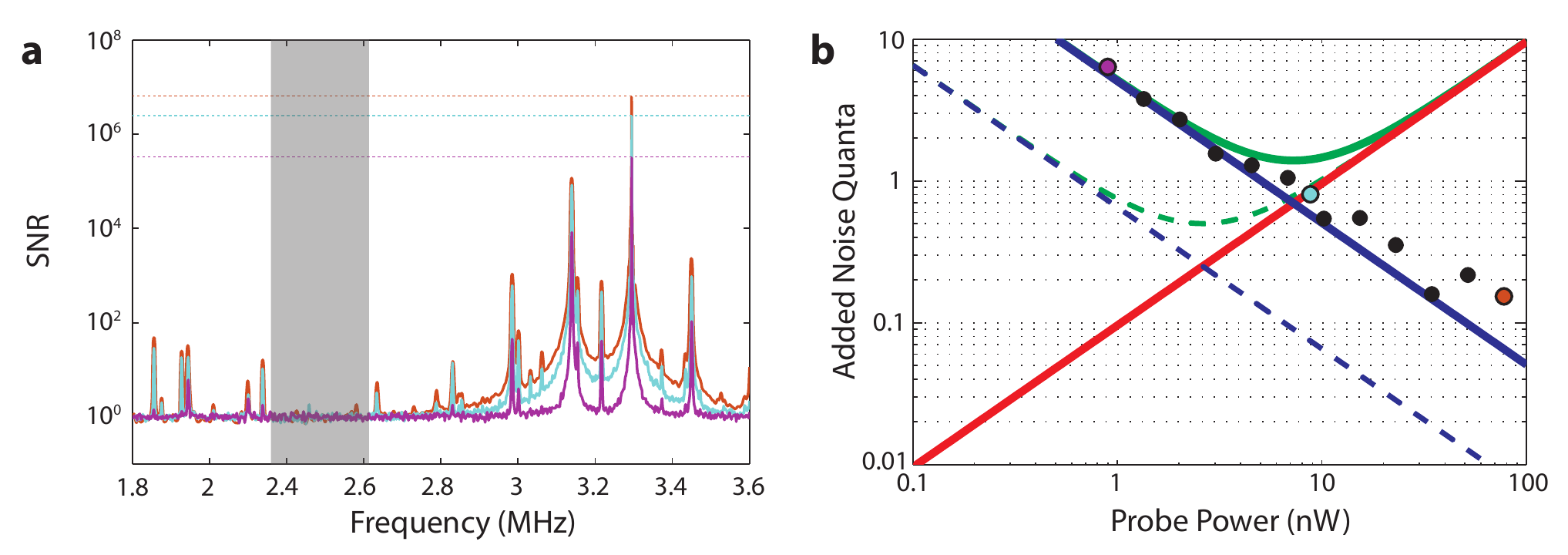}
\caption[width=\columnwidth]{(a) Signal-to-noise ratio (SNR) of $S_{II}(\omega)$ measured with $880$ pW (magenta), $8.6$ nW (cyan), and $76$ nW (orange) probe power from which the imprecision point of corresponding color is extracted in (b). The dashed peak levels are referenced to the background level indicated by the gray shaded region. (b)  Noise quanta versus probe power. Measurements of imprecision are plotted with calculated imprecision (blue), estimated back-action (red), and total (green) noise quanta are plotted in solid curves for the device under test and dashed curves for an ideal measurement.}
\label{fig:fig_5}
\end{figure*}

We determine the imprecision in units of quanta for a measurement of this mechanical mode by referencing the measured background level to the height of the measured noise peak. That is, the number of imprecision quanta is equal to $(\langle n \rangle+1/2)$ divided by the signal-to-noise ratio (SNR) of the resolved mechanical noise peak to the imprecision noise floor, so that an imprecision level of $N_{\text{imp}} = 1$ corresponds to the equivalent level of NPSD that would be produced at the mechanical mode noise peak by a single quantum in the mechanical resonator~\cite{Chan2011}. To determine the background noise floor we take the average of $S_{II}(\omega)$ from $2.4-2.6$ MHz, indicated by relatively noise-free spectral region shaded gray in Fig. \ref{fig:fig_5}a.

To directly measure the additional phonon occupation due to back-action it is necessary to determine the total phonon occupancy, $\langle n \rangle$, by integrating over the full bandwidth of the mechanical mode. However, such a study is outside the scope of the work presented here, as the large thermal occupation at room temperature ($N_{\text{th}} \approx 10^6$) dominates the signal over the comparatively small back-action, $N_{\text{BA}}$, produced by reasonable laser probe powers. For now we restrict our attention to measuring the imprecision, and assume that the backaction is due only to the ideal radiation pressure term given in Eqn. \ref{eqn:Noise}. This assumption is supported by measurements of the properties of known noise sources such as technical laser noise~\footnote{While in principle both intensity and phase noise of the laser can contribute to the heating of the mechanical mode, for an optically resonant measurement of position in the sideband unresolved regime the phase noise does not contribute to backaction and only intensity noise affects the mechanics. In this regime the phase noise adds a small component to the imprecision noise floor. Measurements of the phase and intensity noise of our laser reveal no excess intensity noise and a flat frequency NPSD of $S_{\omega \omega} = 5\times 10^3$ rad$^2$ Hz in the frequency range of interest. Consequently, for the probe powers used in this work, the excess back-action due to technical laser noise is negligible, and the phase noise contribution to the noise floor lies roughly $60$ dB below the shot noise.}, which is a common concern in optomechanical systems~\cite{Rabl2009,Safavi2012,Jayich2012}. As there may exist additional, unknown sources of excess back-action in our measurement, we emphasize that the main result of this work is the collection efficiency of the coupling scheme, and that back-action and noise-driven occupation levels are used only to compare the ideal quantum limits of the device to the SQL.

By measuring the spectrum exemplified in Fig \ref{fig:fig_4} for a range of probe powers we observe the power dependence of $N_{\text{imp}}$, plotted in Fig \ref{fig:fig_5}b with calculations of imprecision (solid blue), ideal quantum back-action (solid red), and total (solid green) noise quanta. Losses in the optical circuit set $\eta_{\text{meas}}=36\%$, which combined with $\eta_{\text{CE}}=52\%$ and electronic noise brings the imprecision level of the measurement to $2.8$ times the ideal imprecision of a lossless device (dashed blue curve). The predicted total minimum added noise is thus $N_{\text{min}}=1.4$ phonons. This minimum total added noise can be compared to similar measurements of mechanical position performed in the optical~\cite{Anetsberger2010,Purdy2012} and microwave~\cite{Teufel2009} domains. The coupling scheme here compares favorably to the minimum total added noise of $N_{\text{min}} \approx 3.2$ demonstrated with whispering-gallery-mode resonators~\cite{Anetsberger2010}, $N_{\text{min}} \approx 0.82$ demonstrated with membranes in Fabry-Perot cavities \cite{Purdy2012}, and $N_{\text{min}} \approx 1.2$ demonstrated with microwave resonators~\cite{Teufel2009}.

Alternatively, one can compare to state-of-the-art ground state cooling experiments~\cite{Teufel2011,Chan2011,Riviere2011} where the occupation of the mechanical resonator is brought near or below a single phonon and the quantum limits of the mechanical measurement become crucial. In such experiments the back-action is used to passively cool the mechanical resonator, and thus the relevant noise term is simply $N_{\text{imp}}$. In the red-detuned, sideband resolved regime relevant for ground-state cooling, the imprecision cannot be made arbitrarily small by increasing the power, but rather has the asymptotic form $N_{\text{imp}} = 1/(4 \eta_{\text{CE}} \eta_{\text{meas}})$ in the limit of large intracavity photon number. Our coupling scheme allows us to achieve $N_{\text{imp}} = 1.34$, which lies well below the imprecision level previously demonstrated using similar nanoscale optomechanical cavties ($N_{\text{imp}}=20$) \cite{Chan2011} , and is comparable to the imprecision acheived in whispering-gallery-mode resonators ($N_{\text{imp}}=3.6$) \cite{Riviere2011} and microwave resonators ($N_{\text{imp}}=1.9$) \cite{Teufel2011}.

It is worth noting that improvements to $\eta_{\text{meas}}$ of the optical circuit such as higher efficiency photodiodes can lower $N_{\text{min}}$ to below $1$ quantum. In addition to improving the collection efficiency towards the theoretical maximum of $\eta_{\text{CE}} = 86\%$, further work will focus on measuring the back-action exerted on the resonator by placing the device in suitable cryogenic conditions where the thermal occupation is strongly suppressed. Another key application which benefits greatly from the improved collection efficiency demonstrated here is feedback damping of the mechanical motion~\cite{Genes2008,Kleckner2006}, which is fundamentally limited by the imprecision noise and could enable ground state cooling of sideband unresolved mechanical systems such as the zipper cavity. 

\section{Acknowledgements}
This work was supported by the DARPA/MTO ORCHID and MESO programs, the Institute for Quantum Information and Matter, an NSF Physics Frontiers Center with support of the Gordon and Betty Moore Foundation, and by the AFOSR QuMPASS MURI. We gratefully acknowledge critical support and infrastructure provided for this work by the Kavli Nanoscience Institute at Caltech.

$^{\dag}$These authors contributed equally to this work


\begin{thebibliography}{10}
\newcommand{\enquote}[1]{``#1''}
\expandafter\ifx\csname url\endcsname\relax
  \def\url#1{{#1}}\fi
\expandafter\ifx\csname urlprefix\endcsname\relax\def\urlprefix{}\fi

\bibitem{Eichenfield20092}
M.~Eichenfield, J.~Chan, R.~M. Camacho, K.~J. Vahala, and O.~Painter,
  \enquote{Optomechanical crystals,} Nature {\bf 462}, 78--82 (2009).

\bibitem{Teufel2011}
J.~Teufel, T.~Donner, D.~Li, J.~Harlow, M.~Allman, K.~Cicak, A.~Siroi,
  J.~Whittaker, K.~Lehnert, and R.~Simmonds, \enquote{Sideband cooling of
  micromechanical motion to the quantum ground state,} Nature {\bf 475},
  359--363 (2011).

\bibitem{Chan2011}
J.~Chan, T.~Mayer-Alegre, A.~H. Safavi-Naeini, J.~T. Hill, A.~Krause,
  S.~Gr\"{o}blacher, M.~Aspelmeyer, and O.~Painter, \enquote{Laser cooling of a
  nanomechanical oscillator into its quantum ground state,} Nature {\bf 478},
  89--92 (2011).

\bibitem{Eichenfield2009}
M.~Eichenfield, R.~Camacho, J.~Chan, K.~J. Vahala, and O.~J. Painter,
  \enquote{A picogram- and nanometre-scale photonic-crystal optomechanical
  cavity,} Nature {\bf 459}, 550--555 (2009).

\bibitem{Clerk2010}
A.~A. Clerk, M.~H. Devoret, S.~M. Girvin, F.~Marquardt, and R.~J. Schoelkopf,
  \enquote{Introduction to quantum noise, measurement and amplification,}
  Review of Modern Physics {\bf 82}, 1155--1208 (2010).

\bibitem{Teufel2009}
J.~Teufel, T.~Donner, M.~Castellanos-Beltran, J.~Harlow, and K.~Lehnert,
  \enquote{Nanomechanical motion measured with an imprecision below that at the
  standard quantum limit,} Nature Nanotechnology {\bf 4}, 820--823 (2009).

\bibitem{Anetsberger2010}
G.~Anetsberger, E.~Gavartin, O.~Arcizet, Q.~Unterreithmeier, E.~Weig,
  M.~Gorodetsky, J.~Kotthaus, and T.~Kippenberg, \enquote{Measuring
  nanomechanical motion with an imprecision below the standard quantum limit,}
  Physical Review A {\bf 82}, 061\,804 (2010).

\bibitem{Riviere2011}
R.~Rivi\`{e}re, S.~Del\'{e}glise, S.~Weis, E.~Gavartin, O.~Arcizet,
  A.~Schliesser, and T.~Kippenberg, \enquote{Optomechanical sideband cooling of
  a micromechanical oscillator close to the quantum ground state,} Physical
  Review A {\bf 83}, 063\,835 (2011).

\bibitem{Michael2007}
C.~Michael, M.~Borselli, T.~Johnson, C.~Chrystal, and O.~Painter, \enquote{An
  optical fiber-taper probe for wafer-scale microphotonic device
  characterization,} Optics Express {\bf 15}, 4745--4752 (2007).

\bibitem{Spillane2003}
S.~Spillane, T.~Kippenberg, O.~Painter, and K.~Vahala, \enquote{Ideality in a
  fiber-taper-coupled microresonator system for application to cavity quantum
  electrodynamics,} Physical Review Letters {\bf 91}, 043\,902 (2003).

\bibitem{Mitomi1994}
O.~Mitomi, K.~Kasaya, and H.~Miyazawa, \enquote{Design of a single-mode tapered
  waveguide for low-loss chip-to-fiber coupling,} IEEE Journal of Quantum
  Electronics {\bf 30}, 1787--1793 (1994).

\bibitem{Almeida2003}
V.~R. Almeida, R.~R. Panepucci, and M.~Lipson, \enquote{Nanotaper for compact
  mode conversion,} Optics Letters {\bf 28}, 1302--1304 (2003).

\bibitem{McNab2003}
S.~J. McNab, N.~Moll, and Y.~A. Vlasov, \enquote{Ultra-low loss photonic
  integrated circuit with membrane-type photonic crystal waveguids,} Optics
  Express {\bf 11}, 2927--2939 (2003).

\bibitem{Chen2010}
L.~Chen, C.~R. Doerr, Y.-K. Chen, and T.-Y. Liow, \enquote{Low-loss and
  broadband cantilever couplers between standard cleaved fibers and
  high-index-contrast Si3N4 or Si waveguides,} IEEE Photonics Technology
  Letters {\bf 22}, 1744--1746 (2010).

\bibitem{COMSOL}
COMSOL Multiphysics http://www.comsol.com/ .

\bibitem{SnyderLove}
A.~W. Snyder and J.~D. Love, {\em Optical Waveguide Theory\/} (Springer, 1983).

\bibitem{Lumerical}
Lumerical Solutions Inc. http://www.lumerical.com/tcad-products/fdtd/ .

\bibitem{Lin2009}
Q.~Lin, J.~Rosenberg, X.~Jiang, K.~Vahala, and O.~Painter, \enquote{Mechanical
  Oscillation and Cooling Actuated by the Optical Gradient Force,} Physical
  Review Letters {\bf 103}, 10\,360 (2009).

\bibitem{Rabl2009}
P.~Rabl, C.~Genes, K.~Hammerer, and M.~Aspelmeyer, \enquote{Phase-noise induced
  limitations on cooling and coherent evolution in optomechanical systems,}
  Physical Review A {\bf 80}, 063\,819 (2009).

\bibitem{Safavi2012}
A.~H. Safavi-Naeini, J.~Chan, J.~T. Hill, S.~Gr\"{o}blacher, H.~Miao, Y.~Chen,
  M.~Aspelmeyer, and O.~Painter, \enquote{Laser noise in cavity-optomechanical
  cooling and thermometry,} arXiv:1210.2671v1 .

\bibitem{Jayich2012}
A.~M. Jayich, J.~C. Sankey, K.~Bjorke, D.~Lee, C.~Yang, M.~Underwood,
  L.~Childress, A.~Petrenko, S.~M. Girvin, and J.~G.~E. Harris,
  \enquote{Cryogenic optomechanics with a Si3N4 membrane and classical laser
  noise,} New Journal of Physics {\bf 14}, 115\,018 (2012).

\bibitem{Purdy2012}
T.~P. Purdy, R.~W. Peterson, and C.~A. Regal, \enquote{Observation of radiation
  pressure shot noise,} arXiv:1209.6334v1 .

\bibitem{Genes2008}
C.~Genes, D.~Vitali, P.~Tombesi, S.~Gigan, and M.~Aspelmeyer,
  \enquote{Ground-state cooling of a micromechanical oscillator: Comparing cold
  damping and cavity-assisted cooling schemes,} Physical Review A {\bf 77},
  033\,804 (2008).

\bibitem{Kleckner2006}
K.~D. and D.~Bouwmeester, \enquote{Sub-kelvin optical cooling of a
  micromechanical resonator,} Nature {\bf 444}, 75--78 (2006).

\end{thebibliography}
\end{document}